\documentclass{article}
\usepackage{verbatim}
\pdfoutput=1
\usepackage{color}
\usepackage{graphicx}
\usepackage{times}
\usepackage{amsmath}
\usepackage[bookmarks=false]{hyperref}
\usepackage{amssymb}
\usepackage{array}
\usepackage{cleveref}
\usepackage{hyperref}
\usepackage[english]{babel}
\usepackage[latin1]{inputenc}
\usepackage{times}
\usepackage[T1]{fontenc}

\newcommand{\be}{\begin{equation}}
\newcommand{\ee}{\end{equation}}
\newcommand{\bx}{\begin{xalignat}{2}}
\newcommand{\ex}{\end{xalignat}}
\numberwithin{equation}{section}

\title{Dyons, Superstrings, and Wormholes } 
\author{
Edward A. Olszewski  \\ 
 {\em Department of Physics} \\ 
 {\em University of North Carolina at Wilmington} \\  
 {\em Wilmington, North Carolina 28403-5606} \\
{\em email: olszewski@uncw.edu}
}
\date{} 

\begin{document}

\bibliographystyle{plain}

\maketitle

\begin{abstract}
We construct dyon solutions on a collection of coincident D4-branes, obtained by applying the group of T-duality transformations to a type I $SO(32)$ superstring theory in 10 dimensions.  The dyon solutions, which are exact,  are obtained from an action consisting of the non-abelian Dirac-Born-Infeld action and Wess-Zumino-like action.  When  one of the spatial dimensions of the D4-branes is taken to be vanishingly small, the  dyons are analogous to the 't Hooft/Polyakov monopole residing in a $3+1$ dimensional spacetime, where the component of the Yang-Mills  potential transforming as a Lorentz scalar is re-interpreted as a Higgs boson transforming in the adjoint representation of the gauge group.  We next apply a T-duality transformation to the vanishingly small spatial dimension.  The result is a collection of D3-branes not all of which are coincident.  Two of the D3-branes which are separated from the others acquire intrinsic, finite, curvature and are connected by a wormhole. The dyon possesses electric and magnetic charges whose values on each D3-brane are the negative of one another. The gravitational effects, which arise after the T-duality transformation, occur despite the fact that the Lagrangian density from which the dyon solutions have been obtained does not explicitly include the gravitational interaction. These solutions provide  a simple example of the subtle relationship between the  Yang-Mills and gravitational interactions, i.e.\@ gauge/gravity duality.    

\end{abstract}

\setcounter{page}{1}
\setcounter{section}{0}
\setcounter{equation}{0}
\section{Introduction}
\label{sect1}
Theoretically appealing but experimentally elusive the magnetic monopole has captured the interest of the physics community for more than eight decades. The magnetic monopole (an isolated north or south magnetic pole) is conspicuously absent from the Maxwell theory of electromagnetism.  In 1931 Paul Dirac showed that the magnetic monopole can be consistently incorporated into the Maxwell theory with virtually no modification to the theory~\cite{diracp31}.  In addition, Dirac demonstrated that  the existence of  a single magnetic monopole necessitates not only that electric charge be quantized but also that the electric and magnetic couplings be inversely proportional to each other, the first suggestion of so called weak/strong duality.  Subsequently, Gerardus 't Hooft~\cite{thooftg76} and Alexander Polyakov showed that  within the context of the spontaneously broken Yang-Mills gauge theory $SO(3)$, topological magnetic monopole solutions of finite mass must necessarily exist. Furthermore, these solutions possess an internal structure and also exhibit the same weak/strong duality discovered by Dirac.  Consequently, Claus Montonen and David Olive conjectured that there exists an exact weak/strong, electromagnetic duality for the spontaneously broken $SO(3)$ gauge theory~\cite{montonenc77}. More recently, this conjecture has become credible within the broader context of $N=2$ or $N=4$ Super-Yang-Mills theories. Despite the lack of experimental evidence for the existence of magnetic monopoles physicists still remain optimistic of their existence. Indeed, Alan Guth proposed the inflationary model of the universe, in part, to explain why magnetic monopoles have escaped discovery~\cite{gutha81}.  

The focus of our investigation is  electrically charged magnetic monopole (dyon) solutions within the context of superstring theory.  In Section~\ref{sect2_1} we construct dyon solutions which are exact and closed to first order in the string theory length scale.  We, first, begin with a type I $SO(32)$ string theory in ten dimensions, six of the spatial dimensions being compact but arbitrarily large.  We, then, apply the group of T-duality transformations to five of the compact spatial dimensions to obtain 16 D4-branes, some of which are coincident.    The five T-dualized dimensions of each D4-brane constitute the  internal dimensions  of a $4+1$ dimensional spacetime.  Making an appropriate ansatz we obtain  dyon solutions residing on the D4-branes.   The solutions are  based on an action which  includes coupling of the D4-branes to NS-NS closed strings, the non-abelian Dirac-Born-Infeld action,  and coupling to  R-R closed strings, a Wess-Zumino-like action.
We next apply a T-duality transformation to the D4-branes,   resulting in a collection of D3-branes, some of which are coincident and  two of which are connected by a wormhole. Finally, we interpret the dyon solutions in the context of gauge/gravity duality.

Because of differences in the literature among the systems of units, sign conventions, etc.\@  we present  in Appendix~\ref{sect8} the conventions chosen by us  so that direct comparisons can be made between our results and those of other authors.

\section{Dyons and Dimensional Reduction of Type I $SO(32)$  Theory}
\label{sect2_1}
In this section we construct dyon solutions based on superstring theory.  We begin with the type I $SO(32)$ superstring theory in ten dimensions~\cite{polchinskij298}, six of whose spatial dimensions are compact. Next, we apply the group of T-duality transformations to five of the compact dimensions letting  the size, $R$, of the dimensions become vanishingly small, i.e.\@ $R \rightarrow 0$. These five dimensions are the internal dimensions of spacetime.  Strictly speaking, spacetime consists of 16 D4-branes, bounded by $2^5$ orientifold hyperplanes.  Each of the D4-branes comprises four spatial dimensions, three unbounded and one compact.  In what follows we assume that none of the D-branes are close to the orientifold hypersurfaces.  Thus, the theory describing the closed strings in the vicinity of any of the D4-branes is  type II oriented, rather than type II unoriented.  In this particular case, since we have applied the T-duality transformation to an odd number of dimensions, the closed string theory is the type IIa oriented theory.  Furthermore, each end of an open string  must be attached to a D4-brane, which may be the same D4-brane or two different D4-branes. 
If we assume   that the number of coincident D4-branes is $n$ ($2 \le n \le 16$),  then a $U(n)$ gauge group is associated with the open strings attached to the coincident D4-branes.  Given these prerequisite conditions we now construct dyon solutions which reside on these coincident D4-branes.   These solutions are derived from the D-brane action comprising two parts, the Dirac-Born-Infeld action, $S_{\text{DBI}}$, which couples NS-NS closed strings to the D4-brane and the Wess-Zumino-like action, $S_{\text{WZ}}$, which couples R-R closed strings to the D4-brane.

\subsection{Dyon Solutions on  D4-branes}
\label{sect2_2}
The dyon solutions are obtained from the equations of motion derived from the action, $S$, which describes the coupling of closed string fields to a general Dp-brane (which in our case is $p=4$). The action is~\cite{johnsonc03} 
\be 
S = S_{\text{DBI}} + S_{\text{WZ}} \:,   \label{eq2_1}    
\ee
where
\begin{equation}
\begin{split}
 S_{DBI} = -\tau_p \int_{\mathcal{M}_{p+1}}  \text{STr} \{ e^{-\Phi} [(-1)\text{det} & (G_{AB}+ \\ 
                            & + B_{AB}  + 2\pi \alpha^\prime F_{AB}]^{1/2} \} \label{eq2_2}
\end{split}   
\end{equation}
and
\begin{equation}
S_{WZ} = \mu_p \int_{\mathcal{M}_{p+1}}  \left[  \sum_{p^\prime} C_{(p^\prime+1)} \right]  \wedge \text{Tr}e^{2 \pi \alpha^\prime F + B}   \:. \label{eq2_3}
\end{equation}
Here $\tau_p$ is the physical tension of the Dp-brane, and $\mu_p$ is its R-R charge (See Appendix~\ref{sect7} for a discussion of the relationships among the various string parameters.).

The dyon solutions are based on the following ansatz. The dilaton background, $\Phi$, is constant,
\begin{equation} 
\Phi = \Phi_0  \:. \label{eq2_a}
\end{equation} 
and background field $B$ vanishes,
\begin{equation} 
 B_{AB}=0 \: (A,B = 0 \cdots 4) \:. \label{eq2_b}
\end{equation} 
  The metric $G$ is given by
	\be
	G_{AB} = \tilde{G}_{AB}I_n  \:, \label{eq2_c}
	\ee
	where, for our purposes, $\tilde{G}_{AB}$ is restricted so that $\tilde{G}_{00} = -1$ and $\tilde{G}_{44 =1}$.

	For $p=4$  we can re-express the determinant in \cref{eq2_2} as
	\be
	\begin{split}	
\text{det}	(G_{AB}  + 2\pi \alpha^\prime F_{AB}) = &  \text{det}(\tilde{G}_{AB}) [I_n  +\frac{ 2\pi \alpha^\prime}{2!} F_{AB}F^{AB}   \\
                 & - \frac{( 2\pi \alpha^\prime)^2}{(5-4)!} \:(3^2 \cdot 1^2) \:^*(F \wedge F)_E \:^*(F \wedge F)^{E}]
 \:,   \label{eq2_2d}
\end{split}
	\ee
	where 
	\be
	^*(F \wedge F)_E = \frac{\sqrt{|\text{det} (\tilde{G}_{AB})}|}{4!}  F^{AB}  F^{CD} \:\epsilon_{ABCDE}  \:. \label{eq2_2f}
	\ee
	  See appendix~\ref{sect9}, \cref{eqc_18}, for further details.

	The term $I_n$ is the $n$-dimensional identity matrix. The value of $n$ is the dimension of the group $U(n)$ associated with the gauge fields residing on the D4-branes.
	All R-R potentials vanish, except for the one-form potential $C_{(1)}$, which is a constant background field,
	\be
	C_{(1)} = C_4 dx^4   \:,  \label{eq2_d}
	\ee
	for some constant value $C_4$.  The gauge field, $F$, is obtained from the gauge potential $A, (A=A_E dx^E)$, where  
	\be
	A_\mu = A_\mu(x^i) \:, (\mu = 0 \cdots 3; i = 1, \cdots 3) \:, \label{eq2_e}
	\ee
	and 
	\be
	A_4 = A_4(x^i) \: \label{eq2_f}
	\ee
 Note that the gauge potentials  are static, i.e.\@ they do not depend on time, $x^0$, and also do not depend on the spatial coordinate $x^4$.  The gauge field $F, (F=F_{A B} dx^A \wedge dx^B)$, a Lie algebra-valued two form, is given by 
\be
 F = dA  - i A \wedge A   \:     \label{eq2_g}
\ee
(See Appendix~\ref{sect8}.).
 The components of the potential $A_0$  and $A_4$  are constrained in accordance with  the condition 
\be
A_0 \wedge A_4 =0 \ \label{eq2_h}
\ee
 so that $F_{04} =0$.  To facilitate its interpretation we  express $F_{AB}$  as a five dimensional matrix which is explicitly partitioned into electric and magnetic fields  which reside in four dimensional space time  and an additional component of the magnetic field which resides in the additional space dimension, i.e.\@
\be
F_{AB} = \left(  \begin{array}{ccccc}
         0 & E_1  &  E_2 & E_3 & 0  \\
				-E_1 & 0   &  B_3  & -B_2 & -\mathcal{D}_1 A_4 \\
				-E_2 & -B_3   &  0  & B_1 & -\mathcal{D}_2 A_4 \\	
				-E_3 & B_2   &  -B_1  & 0 & -\mathcal{D}_3 A_4 \\	
				 0 & \mathcal{D}_1 A_4 & \mathcal{D}_2 A_4 & \mathcal{D}_3 A_4 & 0  \end{array} \right) \:.  \label{eq2iii}
\ee
 We are seeking dyon solutions.  Therefore, with  foresight, we make the following assumptions: 
\be
\begin{array}{rcl}
   E_i &\equiv& F^{a}_{0i}T_{a} =  F^{(i)}_{0i}T_{(i)}\\
	 B_i &\equiv& 1/2 \:\epsilon_{i}^{jk} F^{a}_{jk} T_{a} = 1/2 \:\epsilon_{i}^{jk} F^{(i)}_{jk} T_{(i)} \\
	\mathcal{D}_i A_4  &\equiv & (\partial_i A_4  - i    A_i \wedge A_4)^a T_a = F^a_{4i} T_a = F^{(i)}_{4i} T_{(i)} \: \label{eq2_j}
\end{array}
\ee
The parenthetical index $(i)$  indicates that there is no summation of that index; however, if an expression contains two indices $i$ without parentheses, then summation of these two indices is implied.
 Furthermore, each matrix element in \cref{eq2iii} includes a generator of  $U(n)$, e.g.\@  $E_i =E^{(i)}_{i}\:T_{(i)}$.  Because we are seeking dyon solutions we may assume without loss of generality that each  $T_{(i)}$ is a generator  in the fundamental representation of a local $U(1) \times SU(2)$ subgroup of $SU(n)$ (See \cref{eq2_25e}, below.).

The action, \cref{eq2_2}, can be more straightforwardly interpreted from the perspective of four dimensional spacetime. Since the action  does not depend on the coordinate $x^4$, we can trivially eliminate $x^4$ from the action by integrating  the $x^4$ coordinate. As a result of the integration the tension of the D4-brane, $\tau_4$, and the Yang-Mills coupling constant, $g_{D4}$  are replaced by those of the D3-brane, $\tau_3$ and $g_{D3}$  (See \cref{a3} and \cref{a5}.).   Let the size, $R_4$, of the the $x^4$-dimension become vanishingly small, i.e.\@ $R_4 \rightarrow 0$.   Then, the field $A_4$ becomes a Lorentz   scalar  transforming as the adjoint representation of the gauge group, and \cref{eq2_j} gives the covariant derivative of $A_4$.  From the perspective of four spacetime dimensions, $A_4$, assumes the role of a Higgs boson transforming as the adjoint representation of the gauge group. 

Substituting equations~\cref{eq2_a,eq2_b,eq2_c,eq2iii} into \cref{eq2_1} and then integrating  the $x^4$ coordinate we obtain
\be
S_{DBI}=  \int d^{3+1}\xi \:\mathcal{L_{\text{DBI}}}  \:,   \label{eq2_4}
\ee
where 
\be
\begin{split}
\mathcal{L_{\text{DBI}}}= &  -\frac{1}{(2 \pi \alpha^\prime)^2 g^2_{D3}} \text{STr} [ \{  |\text{det}  (G_{AB}   + 2\pi \alpha^\prime F_{AB} \}|^{1/2} ]  \\
 = & -\frac{\sqrt{|\text{det} (\tilde{G}_{ij})|} }{(2 \pi \alpha^\prime)^2 g^2_{D3}} \:\text{STr} \{ \mathcal{L}^\prime \}            \:. \label{eq2_5}
\end{split}
\ee
The function $\mathcal{L}^\prime $ is defined as
\be
\begin{split}
\mathcal{L}^\prime =& \:\{I_n - (2 \pi \alpha^\prime)^2(E \cdot E -B \cdot B - \mathcal{D} A_4 \cdot \mathcal{D} A_4) \\ 
             & + (2 \pi \alpha^\prime)^4  (B \cdot \mathcal{D} A_4 )^2  - (2 \pi \alpha^\prime)^4 (E \cdot B)^2            \\
						& - (2 \pi \alpha^\prime)^4 ( E \times  \mathcal{D} A_4) \cdot ( E \times  \mathcal{D} A_4) \}^{1/2}  \:. \:\label{eq2_5c}
\end{split}
\ee
We have used the fact that $\sqrt{|\text{det} (\tilde{G}_{ij})|} =\sqrt{|\text{det} (\tilde{G}_{AB})|}$.
In \cref{eq2_5} the ordering of the generators of the algebra, $T_a$, correspond with the order of the fields as they appear in the equation, e.g.
\be
\begin{split}
\{B\cdot \mathcal{D}A_4\}^2 = &\{ B\cdot \mathcal{D}A_4\}\{B\cdot \mathcal{D}A_4\}  \\
      & = \{ \tilde{G}^{ij}  B^{(i)}_i T_{(i)}(\mathcal{D}_jA_4)^{(j)} T_{(j)}\} \{ \tilde{G}^{ij}  B^{(i)}_i T_{(i)}(\mathcal{D}_jA_4)^{(j)} T_{(j)}\}   \:. \label{eq2_5b}
\end{split}
\ee
Note that ''STr'' indicates that the trace is calculated symmetrically, i.e.\@  the trace is  symmetrized with respect to all gauge indices~\cite{tseytlina97,johnsonc03}. The implication is that the evaluation of the trace requires  that after the expansion of \cref{eq2_5} in powers of the field strengths, all orderings of the field strengths are included with equal weight, i.e.\@  products   of the $T_a$ are replaced by their symmetrized sum, before the trace is evaluated.    This is discussed in detail in the references~\cite{tseytlina97,johnsonc03}.

In \cref{eq2_5} the dot product and cross product of two 3-vectors, e.g. $E$ and $ \mathcal{D} A_4$, are defined as $E\cdot \mathcal{D} A_4 = \tilde{G}^{ij} E_i   \:\mathcal{D}_j A_4 $ and $(E \times  \mathcal{D} A_4)_i =  \:\varepsilon_{i}^{jk} E_j  \:\mathcal{D}_k A_4$. 

In obtaining \cref{eq2_5} we have re-expressed the dilaton $\Phi$, on a D4-brane, in terms of the  dilaton $\Phi'$, on a D3-brane, the two of which are related  by a T-duality transformation  in the $x^4$-dimension.  Specifically, $\Phi$ and $\Phi'$ are related by $e^{\Phi^\prime} =  \alpha^{\prime 1/2} e^\Phi/R_4$.   The constant dilaton background $\phi_0$ has been incorporated into the physical tension $\tau_p$ (See Appendix~\ref{sect7}.).

Substituting \cref{eq2_b,eq2_d,eq2_c,eq2iii} into \cref{eq2_3} we obtain
\be
S_{WZ}= \frac{\mu_4}{2!} \int_{\mathcal{M}_{5}} \:  C_{(1)} \wedge  \text{Tr} \{2\pi\alpha^\prime F \wedge 2\pi\alpha^\prime F \} \:.   \label{eq2_6}
\ee
Integrating  the $x^4$ coordinate in \cref{eq2_6} we obtain
\be
 S_{WZ} = \int d^{3+1}\xi \:\mathcal{L_{\text{WZ}}}  \:,   \label{eq2_7}
\ee
where
\be
\begin{split}
\mathcal{L_{\text{WZ}}} = & \frac{\theta}{4\pi^2} \text{Tr} \{F \wedge F \} \\
                        =  &\sqrt{|\text{det} (\tilde{G}_{ij})|}  \frac{\theta}{4\pi^2} \:\text{Tr}\{E \cdot B\} \\
											= & \sqrt{|\text{det} (\tilde{G}_{ij})|}  \frac{\theta}{4\pi^2} \:E^{i(i)}  B^{(i)}_i \text{STr}\{ T_{(i)} T_{(i)}\} \\	
												= & \sqrt{|\text{det} (\tilde{G}_{ij})|}  \frac{\theta}{8\pi^2} \:E^{i(i)}  B^{(i)}_i   \:, \label{eq2_8}
\end{split}
\ee
where $E^{i(i)}=F^{0i(i)}T_{(i)}$.
 Here
\be
\theta \equiv \frac{C_4}{2!} \:\frac{2 \pi R_4}{\alpha^{\prime 1/2}}     \:.   \label{eq2_9}
\ee
In obtaining~\cref{eq2_8} we have explicitly evaluated $\mu_4$ using~\cref{a4}. \Cref{eq2_8} is associated with the Witten effect. Witten has demonstrated that adding the term \cref{eq2_8} to the  Lagrangian of Yang-Mills theory does not alter the classical equations of motion but does alter the electric charge quantization condition in the magnetic monopole sector of the theory~\cite{polchinskij298,harveyj96,wittene79}.  In summary the action, $S$, for  the D4-brane  is given by
\be
S= \int d^{3+1}\xi   \:\mathcal{L},   \label{eq2_10}
\ee
where
\be
\mathcal{L} =\mathcal{L_{\text{DBI}}} +  \mathcal{L_{\text{WZ}}}  \:.  \label{eq2_11}
\ee
The equations of motion which are obtained from~\ref{eq2_10} are   
\be
\mathcal{D}_\mu P^{\mu \nu} = 0   \:, \label{eq2_11b}
\ee
where
\be
P^a_{\mu \nu} = \frac{\partial \mathcal{L}}{\partial F^{ \mu \nu a} }  \:. \label{eq2_11c}
\ee
In addition the fields $F^a_{\mu \nu}$ satisfy the Bianchi identity
\be
     \mathcal{D}_{[\alpha} F^a_{\beta \gamma]} = 0 \:. \label{eq2_11d}
\ee

To facilitate the ensuing analysis we  transform  the Lagrangian density, $\mathcal{L}$, to the Hamiltonian density, $\mathcal{H}$, using the Legendre transformation 
\be
\mathcal{H} = \text{STr} \{ P_0 \cdot E - \mathcal{L} \} \label{eq2_12}
\ee
where
\be
\begin{split}
P^{(i)}_{0 i} \equiv \frac{\partial \mathcal{L}}{\partial E^{i(i)}} = &  -\frac{\sqrt{|\text{det} (\tilde{G}_{ij})|} }{(2 \pi \alpha^\prime)^2 g^2_{D3}} \:\text{STr}     \{ \frac{X^{(i)}_i }  {\mathcal{L}^\prime} \}  \\
  & + \sqrt{|\text{det} (\tilde{G}_{ij})|}  \frac{\theta}{4\pi^2} \:B^{(i)}_i \text{STr}\{ T_{(i)} T_{(i)}\} \:, \label{eq2_13}
\end{split}
\ee
where
\be
X^{(i)}_i = E_i T_{(i)} +  (E \cdot B)  T_{(i)} \:B_i + (\mathcal{D}A_4 \times [E \times \mathcal{D}A_4])_i \:T_{(i)} \:. \label{eq2_13b}
\ee
After performing detailed calculations we obtain
\be
\begin{split}
\mathcal{H} = \frac{\sqrt{|\text{det} (\tilde{G}_{ij})|} }{(2 \pi \alpha^\prime)^2 g^2_{D3}} \text{STr} \{\mathcal{H}^\prime \}  \:, \label{eq2_15}
\end{split}
\ee
where
\be
\begin{split}
\mathcal{H}^\prime = & \{I_n  + (2 \pi \alpha^\prime)^2 (P_0 \cdot P_0 + B \cdot B  + \mathcal{D}A_4 \cdot \mathcal{D}A_4 )  \\
                 & + (2 \pi \alpha^\prime)^4 ([B \cdot \mathcal{D}A_4]^2 +  [P_0 \cdot \mathcal{D}A_4]^2 \\ 
								& + \frac{[P_0 \times B]^2 + [\mathcal{D}A_4 \times (P_0 \times B)]^2}{(2 \pi \alpha^\prime)^{-2} \:I_n +\mathcal{D}A_4 \cdot \mathcal{D}A_4}) \}^{1/2} \:. \:\label{eq2_15b}
\end{split}
\ee

The electric field $E^{(i)}_i$ can be expressed as a function of $P^{(i)}_{0 i}$,
\be
E^{(i)}_i = \frac{\partial \mathcal{H}} {\partial P^{(i)}_{0 i}} = \frac{\sqrt{|\text{det} (\tilde{G}_{ij})|} }{(2 \pi \alpha^\prime)^2 g^2_{D3}} \text{STr} \{ \frac{Y^{(i)}_i}{ \mathcal{H}^\prime} \}
     \:.   \label{eq2_16} 
\ee
The term $Y^{(i)}_i$ is given by
\be
\begin{split}
Y^{(i)}_i= & (2 \pi \alpha^\prime)^2 P_{0 i}  \:T_{(i)} + (2 \pi \alpha^\prime)^4 [ \mathcal{D}_iA_4 \:T_{(i)} (P_0 \cdot \mathcal{D}A_4)+ (B \times(P_0 \times B))_i T_{(i)} \\
    &  -\frac{(B \times \mathcal{D}A_4)_i \:T_{(i)} (P_0 \cdot(B \times \mathcal{D}A_4)}{(2 \pi \alpha^\prime)^{-2} \:I_n+ \mathcal{D}A_4 \cdot \mathcal{D}A_4} ] \:. \label{eq2_17}
\end{split}
\ee

We seek dyon solutions which are BPS states, i.e.\@ whose  energy $\mathcal{E}$, ($\mathcal{E} = \int d^3 \xi \:\mathcal{H}$), is a local minimum.  
First, we re-express  $\mathcal{H}$,
\be
\begin{split}
\mathcal{H} = \frac{\sqrt{|\det (\tilde{G}_{ij})|}}{(2 \pi \alpha^\prime)^2 g^2_{D3}} \text{STr} [ & \{[I_n  + (2 \pi \alpha^           \prime)^2 (\cos \phi \:P_0 \cdot \mathcal{D}A_4 + \sin \phi \:B \cdot \mathcal{D}A_4)]^2 \\ 
           &+ (2 \pi \alpha^\prime)^2 [(\sin \phi \:P_0 \cdot  \mathcal{D}A_4 - \cos \phi \:B \cdot \mathcal{D}A_4 )^2  \\ 
					& + (P_0 - \cos \phi \:\mathcal{D}A_4)^2 +  (B -\sin \phi \:\mathcal{D}A_4)^2] \\
								& + (2 \pi \alpha^\prime)^4 \frac{[P_0 \times B]^2 + [\mathcal{D}A_4 \times (P_0 \times B)]^2}{(2 \pi \alpha^\prime)^{-2} \:I_n +\mathcal{D}A_4 \cdot \mathcal{D}A_4}) \}^{1/2}]  \:. \label{eq2_18}
\end{split}
\ee
The mixing angle, $\psi$, between the electric and and magnetic fields of the dyon is defined as
\be
       \tan \psi = \frac{g_m}{g_e}                    \:. \label{eq2_19}
\ee 
The quantities $g_m$ and $g_e$ are the electric and magnetic charges, respectively, of the dyon.
The energy, $\mathcal{E}$, is minimized by constraining the dyon solutions to satisfy
\begin{subequations}
\label{eq2_20}
\begin{align}
P_0 =& \cos \psi \: \mathcal{D}A_4 \\
B = & \sin \psi \: \mathcal{D}A_4 \:. 
\end{align}
\end{subequations}
In \cref{eq2_18} the second and third squared terms are zero as a consequence of  the constraint.  Since $P_0 \propto B$, the fourth squared term is also zero by virtue of 
\be
(P_0 \times B)^k = \frac{P^{(i)}_{0 i} B^{(j)}_{0 j} - P^{(j)}_{0 j} B^{(i)}_{0 i}}{2!}(T_{(i)} T_{(j)} \varepsilon_{ij}^{ \:\:\:k} - if_{ij}^{\:\:(k)}T_{(k)}  \:. \label{eq2_21}
\ee
Thus, $\mathcal{H}$ simplifies so that the energy is
\be
\begin{split}
\mathcal{E} = & \frac{1}{(2 \pi \alpha^\prime)^2 g^2_{D3}} \int d^3 \xi \:\sqrt{\det (|\tilde{G}_{ij})|} \\
&  \:\text{Tr} [  I_n  + (2 \pi \alpha^           \prime)^2 (\cos \phi \:P_0 \cdot \mathcal{D}A_4 + \sin \phi \:B \cdot \mathcal{D}A_4)]  \:.\label{eq2_22}
\end{split}
\ee
Substituting \cref{eq2_20} into eqs.~\ref{eq2_15b} through \ref{eq2_17}  and using \cref{eq2_22} we find
\be
E= P_0   \:.  \:\label{eq2_22b}
\ee
In \cref{eq2_22} there are two terms which contribute to the mass of the system.  The first term within the trace, i.e.\@ $I_n$, corresponds to the volume of each coincident D4-brane (or D3-brane), which is infinite because the  D-branes are not compact.  The second term, by virtue of the equations of motion, \cref{eq2_11c}, and the Bianchi identity, \cref{eq2_11d},  can be expressed as a divergence and is therefore a topological invariant.  The second term corresponds to the mass of the dyon and is proportional $\sqrt{g_e^2+g_m^2}$ as discussed below.

The solutions to \cref{eq2_11b,eq2_11d}  can be straightforwardly obtained from the dyon solutions derived in reference~\cite{olszewskie12}.  Adapting the notation of reference~\cite{olszewskie12} to the notation used here we  express the vector potential $A$, \cref{eq2_e,eq2_f}, in the form\footnote{In accordance with our conventions the Yang-Mills coupling constant appears explicitly in the Lagrangian \cref{eq_b1}. In references~\cite{harveyj96,olszewskie12} the coupling constant has been incorporated into the Yang-Mills fields.  Thus, to compare results here with those in the references the fields $A$ and related fields should be divided by $g_{D3}$.}
\begin{equation}
\begin{split}
A  = & A_\mu dx^\mu + A_4 dx^4 \\
  = &  \cos \psi \:S(r) \:g_{D3}\:v  \:\alpha_1 T_{r} dt \\ 
 & + W(r) [T_{\theta} \sin( \theta) \:n\:d\phi - T_{\phi} d\theta]       
                                  + \:g_{D3}\:v [\alpha_2 T_\perp + Q(r)  \:\alpha_1 T_{r}] dx^4\:,  \:\label{eq2_23}
\end{split}
\end{equation}
where $v$ is an arbitrary constant.  
For the Lie group $SU(n)$ 
\begin{subequations}
\label{eq2_23b}
\begin{align}
\alpha_1 & = \sqrt{\frac{n}{2(n-1)}} \\
\alpha_2 & = -\sqrt{\frac{n-2}{2(n-1)}}
\end{align}
\end{subequations}

Here the $T_i, (i=r, \theta, \phi )$,  constitute a representation of the $SU(2)$ subalgebra and $T_\perp$ commutes with each of the $T_i$. The quantities $r, \theta, \phi$ are the spherical polar coordinates in three dimensions. The elements $T_{r}, T_{\theta}, T_{\phi}$ are related to $T_x, T_y, T_z$,
\begin{subequations}
\label{eq2_25c}
\begin{align}
T_{r}  = & \:T_x \:\sin \theta \cos n_m \phi + T_y  \:\sin \theta \sin n_m \phi 
              + T_z \:\cos \theta  \\
T_{\theta} = & \:T_x \:\cos \theta \cos n_m \phi + T_y \:\:\cos \theta \sin n_m\phi - T_z \:\sin \theta \\
 T_{\phi}  = & \:-T_x \:\sin n_m \phi + T_y \:\cos n_m\phi  \:,   
\end{align}
\end{subequations}
For $SU(n)$  the $n$-dimensional matrices $T_x, T_y, T_z$ and  $T_\perp$ are given by
\begin{subequations}
\label{eq2_25e}
\begin{align}
T_x = & \frac{1}{2} \left(  \begin{array}{ccccc}
         0 & \cdots  &  0 & 0 & 0  \\
				\vdots &  \ddots  &  \vdots &\vdots & \vdots \\
				0 & \cdots  &  0  & 0 & 0 \\	
				0 &  \cdots  &  0  & 0 & 1 \\	
				 0 & \cdots& 0 & 1 & 0  \end{array} \right) \:,  \\
T_y  = & \frac{1}{2} \left(  \begin{array}{ccccc}
         0 & \cdots  &  0 & 0 & 0  \\
				\vdots &  \ddots  &  \vdots &\vdots & \vdots \\
				0 & \cdots  &  0  & 0 & 0 \\	
				0 &  \cdots  &  0  & 0 & -i \\	
				 0 & \cdots& 0 & i & 0  \end{array} \right)  \:, \\ 
T_z = &\frac{1}{2} \left(  \begin{array}{ccccc}
         0 & \cdots  &  0 & 0 & 0  \\
				\vdots &  \ddots  &  \vdots &\vdots & \vdots \\
				0 & \cdots  &  0  & 0 & 0 \\	
				0 &  \cdots  &  0  & 1 & 0 \\	
				 0 & \cdots& 0 & 0 & -1  \end{array} \right) \:, \\
T_\perp = & \frac{1}{2\sqrt{n(n-2)}} \left(  \begin{array}{ccccc}
         -2 & 0  &  \cdots & 0 & 0  \\
				0 &  \ddots  &  \vdots &\vdots & \vdots \\
				\vdots & \cdots  &  -2  & 0 & 0 \\	
				0 &  \cdots  &  0  & n-2 & 0 \\	
				 0 & \cdots& 0 & 0 & n-2  \end{array} \right) \:. 
\end{align}
\end{subequations}
The $T_x, T_y, T_z$ and  $T_\perp$ are suitable linear combinations of specific elements of the Cartan subalgebra of $SU(n)$ (See reference~\cite{olszewskie12} for details.).  The  value of the integer $n_m$ in \cref{eq2_25c} is the integer multiple of the fundamental unit of the dyon's magnetic charge.

These results differ from those of reference~\cite{olszewskie12}.  For a direct comparison, first replace  the  azimuthal angle, $\phi$ in reference~\cite{olszewskie12} with $\phi'$, and extend the domain from $[0, 2 \pi ]$ to  $[0, 2 \pi \:n_m]$, i.e.\@   $\phi' \in [0, 2 \pi \:n_m]$. Now, perform the change of variables $\phi' = n_m \phi$ to the dyon solutions of reference~\cite{olszewskie12} to obtain those given in \cref{eq2_23}.  In addition, apply the same change of variables to the metric in reference~\cite{olszewskie12}  to obtain  the metric $\tilde{G}_{ij}$,
\be
\tilde{G}_{ij} = \left(  \begin{array}{ccc}
         1 & 0  &  0   \\
				0 & r^2   &  0   \\
				0 &  0  &  r^2 n_m^2 \sin^2 \theta    \end{array} \right) \:.  \label{eq2_25d}
\ee
Here  $ r \in [0, \infty]$,   $\theta \in [0, \pi]$, and $\phi \in [0, 2\pi]$.
 This generalizes the results of reference~\cite{olszewskie12} which only applies to dyons with one unit of magnetic charge, i.e.\@ $g_m = \frac{1}{g_{D3}}$. 

The solutions $W(r)$, $Q(r)$ and $S(r)$ are obtained as in reference~\cite{olszewskie12}
\begin{subequations}
\label{eq2_26}
\begin{align}
W(r) & = w(x)  = 1 - \frac{x}{\sinh x}  \\
 Q(r) & = q(x)  = \coth x - \frac{1}{x} \\
 S(r) & = s(x) = q(x) = \coth x - \frac{1}{x}  \:,
\end{align}
\end{subequations}
where 
the dimensionless variable $x$  is related to the radial coordinate $r$,
\be
x= \sin \psi \:g_{D3} \:v \:\alpha_1 \:r  \:. \label{eq2_25}
\ee

The field tensor $F_{AB}$ of a dyon with electric charge $g_e$ and magnetic charge $g_m$,
\be 
g_m = \frac{n_m}{g_{D3}} \:,  \label{eq2_25a}
\ee
can now be obtained from \cref{eq2_23}. Specifically,
\be
\begin{split}
 F_{t r} = &  \:\frac{g_e}{g}  \:S^\prime(r) \:g_{D3} \:v \:\alpha_1  \:T_r             \\
F_{t \theta} = &  [1-W(r)] \:\frac{g_e}{g}  \:S(r) \:g_{D3} \:v \:\alpha_1   \:T_\theta   \\
F_{t \phi} = &  [1-W(r)] \:\frac{g_e}{g}  \:S(r) \:g_{D3} \:v \:\alpha_1 \:n_m \sin \theta\ \:T_\phi  \\
F_{r \theta} = & -W^\prime(r) \:T_\phi    \\
F_{ \phi  r } = &   -W^\prime(r) n_m \sin \theta  \:T_\theta  \\
F_{\theta \phi } = &  -W(r) \: (2-W(r))\: n_m\sin \theta \:T_r   \:,
\end{split}
\label{eq2_27}
\ee
and
\be
\begin{split}
D_r A_4 & =   Q^\prime(r) \:g_{D3} \:v  \: \alpha_1 \:T_r    \:  \\
D_{\theta} A_4 & =  [1-W(r)] \:Q(r) \:g_{D3} \:v \: \:\alpha_1 \:T_\theta  \\
D_{\phi} A_4 & =  [1-W(r)] \:Q(r) \:g_{D3} \:v \: \:\alpha_1 \:n_m \sin \theta \:T_\phi    \:.
\end{split}
\label{eq2_28}
\ee

We now show that gauge invariance of the  action, \cref{eq2_10}, implies $SL(2,Z)$ invariance.  Consider $U(1)$ gauge transformations which are constant at infinity and are also  rotations about the axis $\hat{A_4}=\frac{A_4}{|A_4|}$,  specifically, the gauge transformations~\cite{harveyj96}
\be
\delta A^a_\mu  = \frac{1}{g_{D3} \:v \alpha_1} ( \mathcal{D}_\mu A_4)^a  \:. \label{eq2_30}
\ee
The action \cref{eq2_10} is invariant under these gauge transformations.  According to the Noether method the generator of these gauge transformations, $\mathcal{N}$,  is given by 
\be
\mathcal{N} = \frac{\partial \mathcal{L}}{\partial \partial_0 A^a_\mu}\delta A_\mu^a  \:. \label{eq2_31}
\ee
Substituting the Lagrangian density \cref{eq2_11} into \cref{eq2_31},  we obtain
\be
\mathcal{N} = \frac{\mathcal{G}_e}{g_{D3}} + \frac{g^2_{D3} \:\theta  \:\mathcal{G}_m}{8 \pi^2}  \:, \label{eq2_32}
\ee
where  
\begin{subequations}
\label{eq2_33}
\begin{align}
\mathcal{G}_m & = \frac{1}{v \alpha_1} \int d^3 \xi \:\:\text{STr} \{\mathcal{D}A_4 \cdot B \} \\
\mathcal{G}_e & =    \frac{1}{g^2_{D3} \:v \alpha_1} \int d^3 \xi \:\:\text{Tr}\{\mathcal{D}A_4 \cdot P_0 \}
\end{align}
\end{subequations}
are the magnetic and electric charge operators.
Since rotations of $2\pi$ about the axis $\hat{A_4}$ must yield the identity for physical states, i.e.
\be
e^{2\pi i \mathcal{N}} = 1 \:, \label{eq2_38}
\ee
Applying the $U(1)$ transformation on the left side of \cref{eq2_38} to states in the adjoint representation of $SU(n)$ we find that the eigenstates of $\mathcal{N}$ are quantized with eigenvalue  
\be
  \mathcal{N} = \alpha_1 \:\eta  \:, \label{eq2_38b}
\ee
where $\eta$ is an arbitrary integer.
Substituting \cref{eq2_38b} into \cref{eq2_32} we obtain
\be
g_e =  \:\alpha_1 \:[\eta \:g_{D3} - \frac{ \:\theta' }{2 \pi} \:n_m \:g_{D3}] \:, \label{eq2_34}
\ee
where we have defined $\theta'$ by 
\be
    \theta \equiv \alpha_1 \theta'   \:, \label{eq2_39}
\ee
and used the fact that 
\be
   g_m \:g_{D3} = n_m \:4\pi  \:. \label{eq2_40}
\ee
Taking $\theta'=0$ in \cref{eq2_34} we obtain the the quantization condition for the electric charge
\be
g_e = \eta \:\alpha_1 \:g_{D3}  \label{eq2_29}
\ee
The electromagnetic contribution to the mass (rest energy) of the dyon, $m_{\text{em}}$, can be obtained by substituting \cref{eq2_33} into  \cref{eq2_22} and integrating the second  term within the trace to obtain
\be
m_{\text{em}} = v \alpha_1  \sqrt{g_e^2+g_m^2} \:.     \label{eq2_40b}
\ee

We can now make the $SL(2,Z)$ symmetry explicit.  We first define 
\be
\tau = \frac{\theta'}{2\pi}  + \frac{4\pi }{g^2_{D3}}\:i     \:. \label{eq2_41}
\ee
If $\theta' = 0$, then the weak/strong duality condition $g_{D3} \rightarrow g_m = (4\pi)/g_{D3}$ is equivalent to
\be 
\tau \rightarrow -\frac{1}{\tau}   \:, \label{eq2_42}
\ee
In \cref{eq2_34} the transformation  $\theta' \rightarrow \theta' +2 \pi$ results in identical physical systems with only  states being relabeled.  The transformation  is equivalent to 
\be
\tau \rightarrow \tau + 1 \:. \label{eq2_43}
\ee
Transformations \ref{eq2_42} and \ref{eq2_43} generate the group $SL(2,Z)$. See references~\cite{harveyj96,wittene79} for further details.

 Note that in \cref{eq2_33} $\mathcal{G}_e$ is strictly speaking, not the electric charge  operator because $P_0$ is not the electric field, but rather is its conjugate; however, according to \cref{eq2_16} and \cref{eq2_17} if $\mathcal{D}A_4$ and $B$ become vanishingly small for asymptotically large values of the radial coordinate, then $P_0$ approaches $E$.  Thus, in the asymptotic limit $\mathcal{G}_e$ is the electric charge operator. This distinguishing feature is a direct consequence of the fact that our analysis is based on the Born-Infeld action rather than the Yang-Mills-Higgs action.  In our case this point is inconsequential since $P_0 =E$, exactly.
\subsection{Dyon Solutions on  D3-branes}
\label{sect2_2b}

As emphasized previously    the dyon solutions derived in Section~\ref{sect2_2}, when  interpreted from $3+1$ spactime dimensions, i.e\@ the compactified theory in which \@ $R_4 \rightarrow 0$, are the  't Hooft/Polyakov magnetic monopole or dyon, with the potential $A_4$ being a Higgs boson tranforming in the adjoint representation of the gauge group $U(n)$.
Here our purpose is to reinterpret these dyon solutions in which $R_4 \rightarrow 0$  from the  equivalent T-dual theory. In the T-dual theory the radius $R_4$  is replaced by $R_4'$, ($R_4' = \alpha'/R_4$) so that the radius of the $x^4$-dimension  $R_4' \rightarrow \infty$.  In addition the potential $A_4$ is reinterpreted as the $x^4$-coordinates of the $n$ D3-branes embedded in $4+1$ spacetime. These coordinates can be directly obtained by diagonalizing $A_4$,  \cref{eq2_23}, using a local  gauge transformation which rotates $T_r$ into $T_z$.  The n $x^4$-coordinates are the diagonal elements of the matrix, i.e.\footnote{We are assuming that after the T-duality transformation the D3-branes are far from any orientifold hyperplanes. This can always be accomplished by adding  to the $A_4$ component of the gauge potential a constant $U(1)$ gauge transformation $\theta_0 \:T^0$, $\theta_0$ being a suitable constant  (See Appendix~\ref{sect8}.).}
\be
\begin{split}
A_4 \rightarrow  & \:2 \pi \alpha' \:g_{D3}\:v [\alpha_2 T_\perp + Q(r)  \:\alpha_1 T_{z}]  \\
  & = \:2 \pi \alpha' g_{D3} \:v \left(  \begin{array}{ccccc}
         u_1 & 0  &  \cdots & 0 & 0  \\
				0 &  \ddots  &  \vdots &\vdots & \vdots \\
				\vdots & \cdots  &  u_1  & 0 & 0 \\	
				0 &  \cdots  &  0  & u_2+\tilde{u}(r) & 0 \\	
				 0 & \cdots& 0 & 0 & u_2-\tilde{u}(r)  \end{array} \right) \:,  \label{eq2b_1}
\end{split}
\ee
where 
\begin{subequations}
\label{eq2b_2}
\begin{align}
u_1 = &-\frac{\alpha_2}{\sqrt{n(n-2)}} \:, \\
u_2 = & \frac{\alpha_2}{2} \:\sqrt{\frac{n-2}{n}} \:, \\
\tilde{u}(r)= & \frac{\alpha_1}{2} Q(r) \:. \\
\end{align}
\end{subequations}
Of the n D3-branes $n-2$ of the D3-branes, denoted $\text{D3}_{n-2}$, are coincident.  The $x^4$-coordinate of each is the constant value $2 \pi \alpha' g_{D3} \:v \:u_1 $.  For the remaining two D3-branes, denoted $\text{D3}_1$ and $\text{D3}_2$, the $x^4$-coordinate of each is a function of the radial coordinate $r$. Specifically,  $x^4 = 2 \pi \alpha' g_{D3} \:v \:(u_2 - \tilde{u}(r))$  for   $\text{D3}_1$ and $x^4 =2 \pi \alpha' g_{D3} \:v \:(u_2 + \tilde{u}(r))$ for $\text{D3}_2$, and as a consequence, these two D3-branes have non-vanishing intrinsic curvature.  This occurs despite the fact that before the application of  the T-duality transformation no gravitational interaction is  explicitly present. 
We now introduce the length scale $L_{D3}$ which is  the separation between  $\text{D3}_1$ and $\text{D3}_2$, in the asymptotic limit as the radial coordinate $r \rightarrow \infty$.  It is related to previously defined parameters by   
\be
L_{D3} = \:2 \pi \alpha' \:g_{D3}  \:v \:\alpha_1  \:. \label{eq2b_2b}
\ee
Another relevant length scale is the size of the dyon, i.e.\@ the region of space where all components of the the Yang-Mills field, $F_{AB}$, are non-vanishing. 
According to \cref{eq2_27}, \cref{eq2_26}, and \cref{eq2_25} only the radial components of the electric and magnetic fields are long range, with the remaining components of the fields vanishing exponentially for $x \gg 1$.
 Thus, additional  stucture of the dyon becomes apparent whenever $x \lesssim 1$ or equivalently whenever $r \lesssim 1/(\sin \psi \:g_{D3} \:v \:\alpha_1)$.  We can therefore define the size of dyon $L_d$, as measured from asymptotically flat space, i.e.\@ r $\rightarrow \infty$, to be
\be
L_d = \frac{1}{\sin \psi \:g_{D3} \:v \:\alpha_1} =\frac{2 \pi \alpha'}{\sin \psi \:L_{D3}} \:. \label{eq2b_2c} 
\ee
\begin{figure}[ht]
 \centering
 \includegraphics[height=6cm]{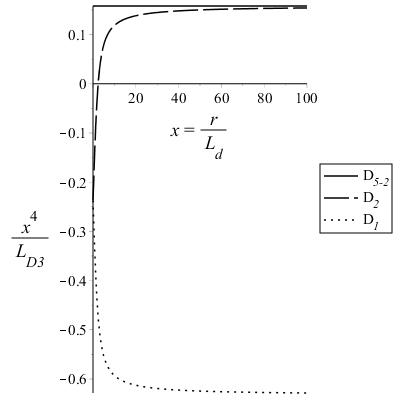}
 \caption{Embedding Functions of the $5$ D3-branes for the Gauge Group $SU(5)$. The scaled coordinate $x^4/L_{D3}$ is plotted as function of the scaled radial coordinate, $r/L_d$, for the $5$ D3-branes. The radial coordinate, $r$, is scaled by the size of the dyon, $L_d$, and the $x^4$ coordinate is scaled by the separation between the two D3-branes $\text{D3}_1$ and $\text{D3}_1$ in the asymptotic limit of large r.}
\label{fig1}
 \end{figure}
In Figure~\ref{fig1}  we show, for the gauge group $SU(5)$, embedding plots of the 5 D3-branes as a function of the dimensionless radial coordinate, $x$, ($x=r/L_d$).  As $x \rightarrow \infty$ the $x^4$-coordinate of  $\text{D3}_2$, approaches that of the the $(5-2)$ coincident D3-branes, $\text{D3}_{5-2}$, in effect, joining them by a wormhole in an asymptotically flat region of space.  At $x=0$,   $\text{D3}_2$ is joined to  $\text{D3}_1$ by a another wormhole. As we shall show, in general, the intrinsic curvature of the two  surfaces in the neighborhood of $x=0$ is relatively large but, nonetheless, finite.  Although these features described in Figure~\ref{fig1} apply for the particular gauge group $SU(5)$, they apply for all $SU(n), \:n \ge 2$.\footnote{For $n=2$ there are no conincident D3-branes.}

We now consider in detail  the two D3-branes $\text{D}_1$ and $\text{D}_2$ whose $x^4$-coordinates are radial dependent.  The T-duality transformation on the $x^4$-dimension pulls back the metric onto $\text{D}_1$ and $\text{D}_2$, inducing the  metric, $G'_{ij}$,
\be
G'_{ij}  = \tilde{G}_{ij}I_n + (\mathcal{D}_i A_4)^{(i)} T_{(i)} \mathcal{D}_j A_4^{(j)} T_{(j)} \delta_{(i) (j)} \tilde{G}_{44} I_n  \:.  \label{eq2b_3}
\ee
Expanding the right-hand side of \cref{eq2b_3} we obtain

\be
 G'_{ij}  =   \left(  \begin{array}{ccccc}
         \tilde{G}_{ij} & 0  &  \cdots & 0 & 0  \\
				0 &  \ddots  &  \vdots &\vdots & \vdots \\
				\vdots & \cdots  &  \tilde{G}_{ij}  & 0 & 0 \\	
				0 &  \cdots  &  0  & \tilde{G}_{ij}+ \tilde{A}_{ij}& 0 \\	
				 0 & \cdots& 0 & 0 & \tilde{G}_{ij}+\tilde{A}_{ij}  \end{array} \right) \:,  \label{eq2b_4}
\ee
where
\be
\begin{split}
\tilde{A}_{ij} = & \left( \frac{g_{D3}\:L \:\alpha_1}{2} \right)^2  \\
                 & \times \left(  \begin{array}{ccc}
         [Q^\prime(r)]^2 & 0  &  0   \\
				0 &  \tilde{W}^2(r)   &  0   \\
				0 &  0  &  \tilde{W}^2(r) \:n^2_m \sin^2 \theta    \end{array} \right) \:.  \label{eq2b_5}
\end{split}
\ee
Here 
\be
\tilde{W}(r) = [1-W(r)] \:Q(r) \:,  \label{eq2b_6}
\ee
and $\tilde{G}_{ij}$ is given by \cref{eq2_25d}.  In obtaining \cref{eq2b_4} we have used \cref{eq2_28} and the fact that the matricies $T_r$, $T_\theta$, and $T_\phi$ ,\cref{eq2_25c}, satisfy the relationship
\be
T^2_r =T^2_\theta =T^2_\phi = \left(\frac{1}{2} \right)^2 \left(  \begin{array}{ccccc}
         0 & \cdots  &  0 & 0 & 0  \\
				\vdots &  \ddots  &  \vdots &\vdots & \vdots \\
				0 & \cdots  &  0  & 0 & 0 \\	
				0 &  \cdots  &  0  & 1 & 0 \\	
				 0 & \cdots& 0 & 0 & 1  \end{array} \right) \:. \label{eq2b_7}
\ee 
Each of the diagonal entries in the matrix $G'_{ij}$ corresponds to the metric on one of the n D3-branes obtained from the T-duality transformation.  In the case of the first $n-2$ entries, corresponding to the D3-branes $\text{D3}_{n-2}$, the metric is flat.   In the case of the last two entries corresponding to the D3-branes, $\text{D3}_1$ and $\text{D3}_2$, their geometries are identical and intrinsically curved. The only feature which distinguishes these two D3-branes is that the function $Q(r)$ defining the $x^4$-coordinate for D3-brane $\text{D}_2$  is replaced by $-Q(r)$ for $\text{D3}_1$, as evidenced in \cref{eq2b_1} and \cref{eq2b_2}, and also in Figure~\ref{fig1}.  As a consequence, the electric and magnetic charge of the dyon on $\text{D}_2$ is minus the values on $\text{D}_1$.  The electric and magnetic field lines enter the wormhole from one D3-brane and exit from the other.
\begin{figure}[ht]
 \centering
 \includegraphics[height=6cm]{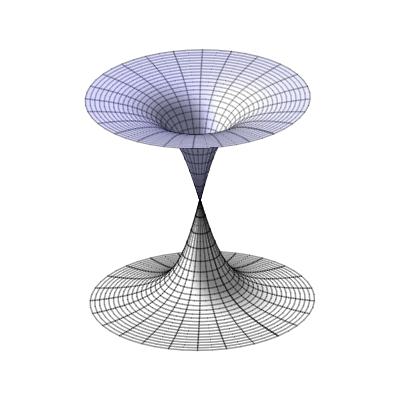}
 \caption{Wormhole.  Shown is the embedding diagram for the two D3-branes, $\text{D3}_1$, $\text{D3}_2 $, with azimuthal angle suppressed.  The domain of the radial coordinate,  $r$, is $0 \le r \le 15 L_d$, and the range of the embedding coordinate $x^4$ is  $-.45 L_{D3} \le x^4 \le +.45 L_{D3}$. }
\label{fig2}
 \end{figure}
Figure~\ref{fig2} is an embedding diagram showing the D3-branes $\text{D}_1$ and $\text{D}_2$ in the neighborhood of the radial coordinate $r=0$.  As there is no event horizon surrounding $r=0$ the two D3-branes are joined by a wormhole at $r=0$. 

Of particular interest is the intrinsic scalar curvature of $\text{D3}_1$ and $\text{D3}_2$, in the neighborhood of $r=0$.  The scalar curvature can be calculated from the metric $G'_{ij}$, \cref{eq2b_4}, and its value $R(0)$  at $r=0$ is
\be
R(0) = 216 \sin \psi  \:\frac{L^6_{D3} \:\sin^3 \psi}{[L^4_{D3} \:\sin^2 \psi  + (12 \pi \alpha')^2]^2   }    \:.  \label{eq2b_8}
\ee
For a given value of $\sin\psi$, $R(0)$ assumes its maximum value $\tilde{R}(0)$,  
\be
  \tilde{R}(0) = 
	\frac{27 \sqrt{3}}{8} \frac{\sin \psi}{ \pi \alpha'}  \:, \label{eq2b_10}
\ee 
when $L_{D3} = \tilde{L}_{D3}$ where
\be
\tilde{L}_{D3} =  \left( \frac{12\sqrt{3} \:\pi \alpha' }{\sin \psi} \right)^{1/2}   \:. \label{eq2b_9}
\ee
For either $L_{D3} \rightarrow 0$ or $L_{D3} \rightarrow \infty$ the scalar curvature $R(0) \rightarrow 0$, i.e.\@ the geometry of $\text{D}_1$ and $\text{D}_2$  becomes flat, everywhere.  The expression for the scalar curvature $R(r)$ is a complicated function of $r$ and not amenable to straightforward interpretation, and therefore, will not be given.  
\begin{figure}[ht]
 \centering
 \includegraphics[height=6cm]{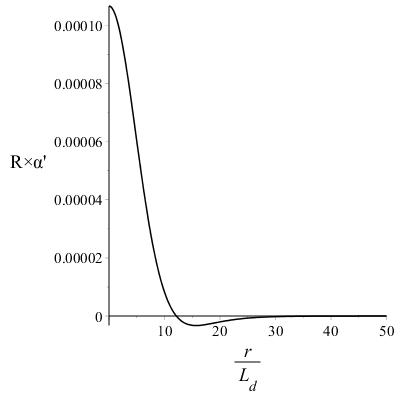}
 \caption{Scalar Curvature $R(r)$. For the case $L_{D3} = \sqrt{\alpha'}$ and $\sin \psi =1$ the dimensionless scalar curvature, $R \times \alpha'$, is plotted as as a function of the scaled radial coordinate, $r/L_d$.    }
\label{fig3}
 \end{figure}
In Figure~\ref{fig3} we show a plot of the scalar curvature as a function of the radial coordinate. In this example $L_{D3} = \sqrt{\alpha'}$, and the dyon  has only one unit of magnetic charge so that $\sin \psi =1$. Near $r=0$ the scalar curvature is positive and finite.  As $r$ increases the scalar curvature becomes slightly negative and asymptotically approaches zero as $r \rightarrow \infty$.   These features of the scalar curvature described for this specific example also apply in general.

 Consider dyon solutions for which $L_{D3} \approx \sqrt{\alpha'}$ or less.  The F-strings connecting $\text{D3}_1$ and $\text{D3}_2$ would be in their ground state, a BPS state.  In addition, assume that $g_{D3} \ll 1$, then as $ r \rightarrow 0$ from an asymptotically flat region of space, within either $\text{D3}_1$ or $\text{D3}_2$, the string length scale will be reached before the gravitational interaction  becomes dominant at the length scale of $\scriptstyle{\mathcal{O}}$$(g^{1/4}_{D3} \:\sqrt{\alpha'})$~\cite{polchinskij298}. Thus, the action, \cref{eq2_2}, which does not include the gravitational interaction, should apply, and consequently, the dyon solutions derived should be accurate.  On the other hand, let $g_{D3} \rightarrow 1/g_{D3}$ so $L_{D3} \gg 1$ then the D-string, also a BPS state, becomes lighter than the F-string.   As a consequence of weak/strong duality the dyon solutions should still be applicable  with the F-strings being replaced by D-strings, and the dyon electric and magnetic charges being interchanged.

After applying a T-duality transformation to the dyon solutions obtained in  Section~\ref{sect2_2} we have obtained dyon solutions residing on D3-branes where the effect of the gravitational interaction is apparent.  This occurs  despite the fact that the action,~\cref{eq2_2}, does not  explicitly include the gravitational interaction. The presence of gravitational effects in this case is an example of how, in string theory, one loop open string interactions, i.e.\@ Yang-Mills interactions, are related to tree level, closed string interactions, i.e.\@ gravitational interactions. 
\begin{figure}[ht]
 \centering
 \includegraphics[height=6cm]{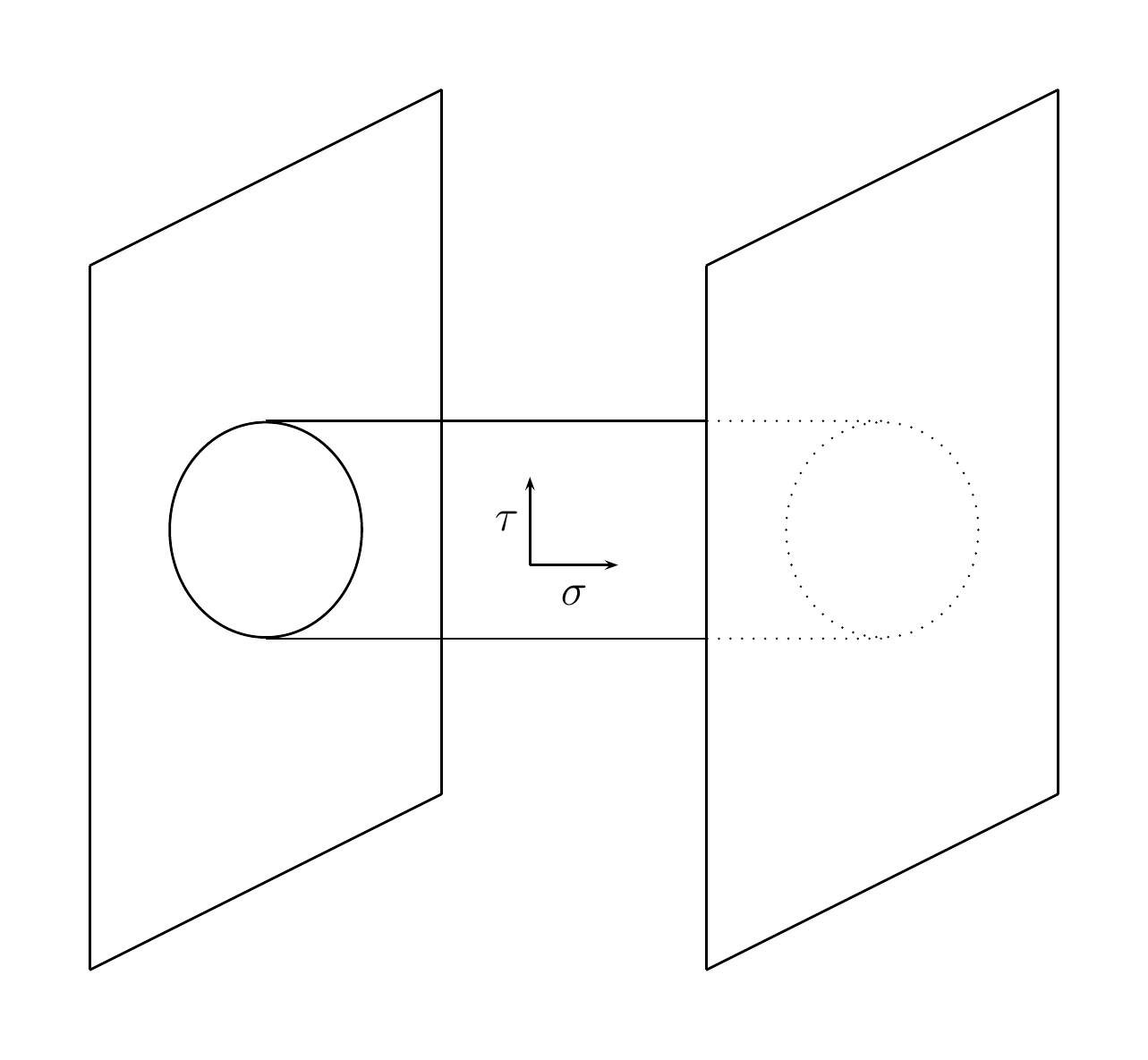}
 \caption{One Loop String Diagram.  Shown is the diagram for an open string, whose end points are fixed on two different Dp-branes, propagating in a loop, $\tau$ being the time variable on the world sheet.  Alternatively, interchanging $\tau$ with $\sigma$ the same diagram can be interpreted as a closed string being exchanged between the two Dp-branes. }
\label{fig4}
 \end{figure}
In Figure~\ref{fig4} we depict, for illustrative purposes, two parallel Dp-branes in close proximity.  The two Dp-branes can interact through open strings which connect the two Dp-branes.  In the Figure we show the one loop vacuum graph for such an interaction which can be interpreted as an open string moving in a loop.   Alternatively, the interaction can be interpretated as a closed string being exchanged between two Dp-branes.  In a certain sense  spin-2 gravitons, i.e.\@ the closed strings in their massless state, comprise a bound state of spin-one Yang-Mills bosons, i.e.\@ open strings in their massless state.   

Prior to the application of the T-duality transformation the open strings can propagate anywhere in the D4-brane.  After the T-duality transformation the open strings are constrained to propagate only within D3-branes; whereas, the closed strings can still propagate in the bulk region between the D3-branes, i.e.\@ gravitons can propagate in spacetime dimensions not allowed for the Yang-Mills bosons.
Based on certain general assumptions Weinberg and  Witten have shown the impossibility of constructing a spin-2 graviton as a bound state of spin-1 gauge fields~\cite{weinbergs80}. One of the assumptions on which the proof is based is that the spin-1 gauge bosons and spin-2 gravitons propagate in the same spacetime dimensions.  In the example presented here,  this assumption is violated so that the conclusion of their theorem is avoided.  These dyon solution, thus, provide a simple example of gauge/gravity duality, which is discussed in detail in the work of Joseph Polchinski~\cite{polchinskij10}.

\setcounter{equation}{0}
\section{Conclusions}
\label{sect6}
We have investigated  dyon solutions within the context of superstring theory.  Beginning with the type I $SO(32)$ superstring theory in ten dimensions, six of whose spatial dimensions are compact, we have applied the group of T-duality transformations to five of the compact dimensions.  The result is 16 D4-branes, a number $n$, ($2 \le n \le 16$), of which are coincident.  The five T-dualized dimensions, whose size is taken to be vanishingly small, become  the five internal spacetime dimensions while the remaining five dimensions  correspond to the external $4+1$ dimensional spacetime. Making a suitable ansatz for the gauge fields residing on the $n$ coincident D4-branes we have obtained dyon solutions  from an   action  consisting of two terms: the $4+1$ dimensional,   non-abelian Dirac-Born-Infeld action and a  Wess-Zumino-like action. The former action gives the low energy effective coupling of  D4 branes to NS-NS closed strings, and the latter  of  D4-branes to R-R closed strings.  The method of solution involves transforming the $4+1$ dimensional action from the Lagrangian formalism to the Hamiltonian formalism and then seeking solutions which minimize the energy. 
The resulting dyon solutions, which are BPS states, reside on the $n$ D4-branes and are therefore associated with a supersymmetric $U(n)$ gauge theory in $4+1$ spacetime dimensions.  These dyon solutions can be alternatively understood in the limit when the size of the remaining compact spacetime dimension, $x^4$, approaches zero.  In this situation the $4+1$ dimensional spacetime is reduced to a $3+1$ dimensional spacetime.  As a consequence, the $A_4$ component of the vector potential becomes a Lorentz scalar with respect to $3+1$ dimensional spacetime, and can be  interpreted as a Higgs boson transforming as the adjoint representation of the $U(n)$ gauge group, analogous to the Higgs boson associated with the\@ 't Hooft/Polyakov  magnetic monopole.  Finally, we perform a T-duality transformation in the $x^4$-direction.  As a result   $n-2$ of the  D4-branes  are transformed into $n-2$ coincident D3-branes, whose intrinsic geometry is flat.  The remaining two D4-branes are transformed into two separate D3-branes whose intrinsic geometry is curved.   As depicted in Figure~\ref{fig3} the two D3-branes are joined by wormhole  at $r=0$. The scalar curvature of each D3-brane  reaches a maximum, finite  value,  at $r=0$  and approaches zero as $r \rightarrow \infty$. The dyon resides on these two D3-branes.   Furthermore, the values of electric and magnetic charges of the dyon on one D3-brane are minus the values on the other D3-brane, and as a consequence the electric and magnetic field lines enter the wormhole from one D3-brane and exit from the other D3-brane.  

The T-duality transformation in the $x^4$-direction causes two of the D3-branes to aquire intrinsic curvature.  This occurs despite the fact that the Lagrangian density from which the dyon solutions have been obtained does not explicitly include the gravitational interaction.  This can be understood heuristically from the open string, one loop vacuum graph given in Figure~\ref{fig4}.  From one perspective the graph describes an open string, whose ends are fixed on two different D3-branes, moving in a loop, or alternatively,  the exchange of a closed string between two D3-branes.  Thus, the gravitational interaction, i.e.\@ the closed string interaction, and  the Yang-Mills interaction, i.e.\@ the open string interaction, appear as alternative descriptions of the same interaction.  This simple example is suggestive of the subtle, but profound, connection between the  Yang-Mills and  gravitational interactions, specifically  gauge/gravity duality.

\appendix
\section{Appendix}
\label{sect8}
Concerning conventions the Minkowski signature is $(-+++ \dots)$, and the Levi-Civit\`{a} symbols $\epsilon_{0123} = \epsilon_{123} = 1$.  
Other relevant conventions are: Greek letters denote four dimensional spacetime indices, i.e.\@ 0, 1, 2, 3, whereas capitalized Roman letters are used when the   spacetime  dimension is greater than four. The small Roman letters, $i, j, k, l, m,$ are reserved for spatial dimensions in four dimensional spacetime, i.e.\@ 1, 2, 3. The small Roman letters $a, b, c, d,$ are used to enumerate the generators of the Lie group. The Levi-Civit\`{a} tensor in three space dimensions is $\varepsilon_{ijk} = \sqrt{|\text{det} (\tilde{G}_{ij})|} \epsilon_{ijk}$, where the $\tilde{G}_{ij}$ are the spatial components of the the metric tensor.  We focus our attention on Yang-Mills theories based on the compact Lie groups, $U(n)$. A typical group element $u, (u \in U(n))$ is represented in terms of the group parameters $\theta_a, (a= 0 \cdots n^2-1)$ as $u =  e^{i \theta_a T^a}$. The generators  of this group in the fundamental representation are denoted $T^a, (a= 0 \cdots n^2-1)$. The generator  $T^0$ generates the $U(1)$ portion of $U(n)$, and the remaining $T^a$ generate the $SU(n)$ portion. The Lie algebra of the group generators, $T^a$, is $[T^b, T^c ]= i f^{abc} T^c$, the $f^{abc}$ being the structure constants of $U(n)$. The generators of the $U(n)$  are required to  satisfy the trace condition $\text{Tr} (T^a T^b) = \delta^{ab}/2$. Thus, in particular, $T^0 = (1/\sqrt{2n}) I_n, I_n$ being the $n$-dimensional identity matrix.  The Yang--Mills coupling constant is denoted $g_{D3}^2$.  We  employ Lorentz-Heaviside units of electromagnetism  so that $c=\hbar=\epsilon_0=\mu_0=1$; Consequently, the Dirac quantization condition  is $g_{D3} \:g_m  = 2 \pi$. The quantity $g_m$ is the magnetic charge of a  unit charged  Dirac monopole.

Consistent with our analysis the Yang-Mills-Higgs Lagrangian is 
 \be
\begin{split}
\mathcal{L} & = -\frac{1}{2 g^2_{D3}} \text{Tr} (F_{\mu \nu}  F^{\mu \nu} ) +  \text{Tr} (\mathcal{D}_\mu W  \mathcal{D}^\mu W )- V(2\text{Tr}[WW])   \\  
              & =   -\frac{1}{4 g^2_{D3}}  F^a_{\mu \nu}  F^{a \mu \nu}  + \frac{1}{2}  \mathcal{D}_\mu W^a  \mathcal{D}^\mu W^a - V(W^a \:W^a)   \:,\label{eq_b1}
\end{split}
\ee
where $V$ is a potential function depending on the Higgs field, $W$, and
\be
\begin{array}{rcl}
   F_{\mu \nu} &=& F^a_{\mu \nu} T^a   \\
	  W           &=& W^a T^a     \\
		A_\mu       &=&A^a_\mu T^a   \:.\label{eq_b2}
\end{array}
\ee
The covarient derivitative, $\mathcal{D}_\mu$, is defined as
\be
\mathcal{D}_\mu  \equiv \partial_\mu - i A_\mu \:, \label{eq_b3}
\ee
\be
        F_{\mu \nu}  = -i [\mathcal{D}_\mu, \mathcal{D}_\nu]  \:. \label{eq_b4}			             
\ee
Thus,
\be
     F^a_{\mu \nu}= \partial_\mu A^a_\nu - \partial_\nu A^a_\mu + f^{abc} A^b_\mu A^c_\nu  \:. \label{eq_b_5}
\ee
The Higgs field $W$ is a scalar transforming as  the adjoint representation of $U(n)$ so that its covariant deriviative is
\be
\begin{split}
\mathcal{D}_\mu W  &=  \partial_\mu W  - i \: [A_{\mu}, W]  \\
        &  = \partial_\mu W^a  + f^{abc} A^b_\mu W^c  \:.\label{eq_b6}
\end{split}
\ee

\section{Appendix}
\label{sect7}
In principle string theory has no adjustable parameters other than its characteristic length scale, $\sqrt{\alpha^\prime}$; however,  various parameters of the theory do depend on values of the background fields.  For reference we provide explicit relationships between various string theory parameters and $\alpha^\prime$.  Let $\Phi_0$ be the  vacuum expectation value of the dilaton background, i.e.\@ $\phi_0=<\Phi>$, $\Phi$ being the dilaton background. The closed string coupling constant, $g$, is
\be
   g = e^{\Phi_0}  \:.  \label{a1}
\ee 
The physical gravitational coupling, $\kappa$, is  
\be
\kappa^2 \equiv \kappa^2_{10} g^2 = 8 \pi G_\text{N}= \frac{1}{2}(2 \pi)^7  \alpha^{\prime 4} g^2 \:,  \label{a2}
\ee
where $G_\text{N}$ is Newton's gravitational constant in 10 dimensions and $\kappa^2_{10} = \kappa^2_{11}/2 \pi R$.  The quantity $\kappa_{11}$  is the gravitational constant appearing in the eleven dimensional, low energy effective action of supergravity, and $R$ is the compactification radius for reducing the the eleven dimensional theory to ten dimensions.  The physical Dp-brane tension, $\tau_p$, is 
\be
  \tau_p \equiv T_p/g =\frac{1}{g (2 \pi)^p \alpha^{\prime (p+1)/2 }}= (2 \kappa^2)^{-1/2} (2 \pi)^{(7-2p)/2} \alpha^{\prime 4}\:,  \label{a3}
\ee
where $T_p$ is Dp-brane tension.  The Dp-brane R-R charge, $\mu_p$, is
\be
\mu_p = g \tau_p = \frac{1}{ (2 \pi)^p \alpha^{\prime (p+1)/2 }} \:.  \label{a4}
\ee
The  coupling constant $g_{Dp}$ of the $U(n)$ Yang-Mills theory on a Dp-brane is given by
\be
 g^2_{Dp} = \frac{1}{(2 \pi \alpha^\prime)^2 \tau_p} = (2 \pi)^{p-2} g \:\alpha^{\prime (p-3)/2 } \:. \label{a5}
\ee
 The ratio of the F-string tension, $\tau_{F1}$, to the D-string (D1-brane) tension, $\tau_{D1}$, is
\be
\frac{\tau_{F1}}{\tau_{D1}} = g  \:.  \label{a6}
\ee 
\section{Appendix}
\label{sect9}
In this Appendix we provide a heuristic derivation of the formula for evaluating an  $N^\prime$-dimensional, ($N^\prime = p+1$), determinent of the form $\text{det} (\tilde{G}_{AB}  + 2\pi \alpha^\prime F_{AB})$, $(A,B= 0 \dots p)$, \cref{eq2_2} and \cref{eq2_b}.\footnote{The notation used in this Appendix does not adhere strictly to the font conventions defined in Appendix~\ref{sect8}.} Without loss of generality we assume that the metric is diagonal.  Consequently, we can express the determinent in the generic form 
\be
\text{det}(h) \epsilon_{I J \cdots O P}= \epsilon_{ij \cdots o p} h_{iI} h_{j J} \cdots h_{o O} h_{p P}  \:, \label{eqc_1}
\ee
where $h_{ij} = f_{ij}=2\pi \alpha^\prime \:F_{(i-1)(j-1)}, (i,j = 1 \cdots N^\prime)$ and $h_{(i)i}= g_{(i)i}=\tilde{G}_{(i-1)(i-1)}$.  Note:  indices enclosed within parentheses are not summed.

Thus, the right-hand side of \cref{eqc_1} comprises a sum of terms, each of which consists of products  of   the metric, $g_{(i)i}$ or $f_{ij}$. We need only consider  terms in which the number of $f_{ij}$ in each product is even, since terms containing products of an odd number of $f_{ij}$ vanish because $F_{AB} = -F_{BA}$. Thus, $\text{det}(h_{ij})$, \cref{eqc_1}, can be re-expressed as a sum 
\be
\text{det}(h) \epsilon_{I J \cdots O P} = (H_0 + H_2 + H_4 + H_{2n^\prime}+ \cdots)   \epsilon_{I J \cdots O P} \:. \label{eqc_1b}
\ee
Each $H_{2n^\prime}$  is a term in \cref{eqc_1} which contains a product of  $f_{ij}$ which is even in number. The values of $2n^\prime$ range from 0 to $N^\prime$ or  $N^\prime-1$ depending on whether  $N^\prime$ is even or odd.
The value of $H_0$, which contains no off-diagonal elements, $f_{ij}$,  evaluates to 
\be
 H_0= \det (g) \equiv g_{1 1} g_{2 2} \cdots g_{N^\prime N^\prime }    \:. \label{eqc_2} 
\ee

In order to understand the structure  of $H_{2n^\prime}$, for arbitrary $n^\prime$, we first  study the structure of $H_4$,  
\be
H_4  \:\epsilon_{I J \cdots O P} = \epsilon_{I J \cdots k l m n \cdots O P} \:g_{(I)I} \:g_{(J)J}   \cdots g_{(O)O} \:g_{(P)P} \:f_{k K}  f_{l L} f_{m M} f_{n N}        \:. \label{eqc_3}
\ee
\Cref{eqc_3} represents a term in \cref{eqc_1} where the values of $(k,l,m, n)$ in the sum are restricted to the specific integer values  $(K,L,M,N)$  from the set of integers $(1, 2, \cdots N^\prime)$. Multiplying \cref{eqc_3} by the Levi-Civit\`{a} symbol (which does not change  value of the expression) we obtain
\be
\begin{split}
H_4 = & g_{(I)I} \:g_{(J)J}   \cdots g_{(O)O} \:g_{(P)P} \\
       & \epsilon_{I J \cdots klmn \cdots O P}  \:f_{k K}  f_{l L} f_{m M} f_{n N}         
\:\epsilon_{I J \cdots KLMN \cdots O P}
        \:. \label{eqc_4}
\end{split}
\ee
We now re-arrange the terms in \cref{eqc_4} in a form which is more useful for the subsequent analysis. By virtue of the Levi-Civit\`{a} symbols none of the $(k,l,m,n)$ is equal to any of the others, and similarly for the $(K,L,M,N)$; however, each of the  
 $(k,l,m,n)$ is equal to one of the $(K,L,M,N)$.  Because of the anti-symmetry of the $f_{ij}$, some of the terms in the sum vanish, i.e. whenever $k = K$,  or  whenever $l = L$, etc. By explicit construction or by a combinitorics argument we can show that there are  nine terms which are non-vanishing.  We consider one typical term  in the sum, e.g.\@  the term, $\{1\}$,
\be 
\{1\}= \{m=K, k = L, n=M, l=N\}  \:. \label{eqc_4b}
\ee
We now show how to re-express \cref{eqc_4} so the half, i.e.\@ 2, of the four $f_{ij}$, are associated with one of the Levi-Civit\`{a} symbols and the other half are associated with the other Levi-Civit\`{a} symbol.  In order for the two $f_{ij}$ to be associated with one Levi-Civit\`{a} symbol, the four subscripts on the two $f_{ij}$ must be different.  We associate the first $f_{kK}$ with the Levi-Civit\`{a} symbol to its left. To determine the remaining associations we proceed as follows.  Since $K=m$, we associate $f_{mM}$ with the Levi-Civit\`{a} symbol to the right.  We now consider the second term in \cref{eqc_4}, i.e.\@ $f_{lL}$.  Since $L = k$, we associate $f_{lL}$ with the  Levi-Civit\`{a} symbol to its right. and $f_{nN}$ with the Levi-Civit\`{a} symbol to the right.  We continue in this manner to the next remaining term, $f_{nN}$.  Since $N$ does not equal any of the lower case Roman subscripts associated with the 
 Levi-Civit\`{a} symbol to the left ($N=l$), we assign $f_{nN}$ to the Levi-Civit\`{a} symbol to the left.  This completes the process   since each $f_{ij}$ is associated with either of the two Levi-Civit\`{a} symbols.  Using  \cref{eqc_4b} we re-express the subscripts on the Levi-Civit\`{a} symbols in \cref{eqc_4}
\be
\begin{split}
H_{4,\{1\}} = & g_{(I)I} \:g_{(J)J}   \cdots g_{(O)O} \:g_{(P)P} \\
       & \epsilon_{I J \cdots (k)NK(n) \cdots O P}  \:f_{k K} f_{n N} f_{m M} f_{l L}            
\:\epsilon_{I J \cdots (m)LM(l) \cdots O P}
        \:. \label{eqc_5}
\end{split}
\ee 

Now, we permute the subscripts of the Levi-Civit\`{a} symbols so that the corresponding lower case and upper case Roman letters are adjoining.  Both the upper case M and K require the same number of movements as do the upper case $N$ and $L$.  Since the number of permutations is even,  no change in sign of the Levi-Civit\`{a} symbols results from permuting the subscripts. \Cref{eqc_5} becomes
\be
\begin{split}
H_{4,\{1\}} = & g_{(I)I} \:g_{(J)J}   \cdots g_{(O)O} \:g_{(P)P} \\
       & \epsilon_{I J \cdots (k)(K)(n)(N) \cdots O P}  \:f_{k K} f_{n N} f_{m M} f_{l L}            
\:\epsilon_{I J \cdots (m)(M)(l)(L) \cdots O P} 
        \:. \label{eqc_6}
\end{split}
\ee
Each of the remaining $H_{4,\{s\}}, (s=2 \cdots 9)$ can be expressed, similarly.  

In order to understand, in generic terms, the structure of $H_4$ consider the following expression $H^\prime_4$,
\be
\begin{split}
H^\prime_{4}= & g_{(I)I} \:g_{(J)J}   \cdots g_{(O)O} \:g_{(P)P} \\
       & \epsilon_{I J \cdots kKnN \cdots O P}  \:f_{k K} f_{n N} f_{m M} f_{l L}            
\:\epsilon_{I J \cdots mMlL \cdots O P} 
        \:. \label{eqc_7}
\end{split}
\ee
The expression $H'_4$  differs from the individual term $H_{4,\{1\}} $ in that the repeated indices $(k,K,n,N) $ and $(m,M,l,L)$ are summed. By inspection the term $H_{4,\{1\}} $, as well as the remaining  terms $H_{4,\{s\}}, (s=2 \cdots 9)$, is contained in $H'_4$.  Overtly, the number of non-vanishing terms in \cref{eqc_7} is $(4!)^2$, each factor of $(4!)$ coming from each one of the Levi-Civit\`{a} symbols.  By a combinatorics arguement each factor of $4!$ is eight-fold redundant so that the number of independent terms associated with each Levi-Civit\`{a} symbol is three.  Consequently, the number of independent terms in $H^\prime_{4}$ is 9, i.e.\@ $3 \times 3$, with a redundancy $8 \times 8$.  Thus,
\be
H_4 =\frac{1}{8 \times 8} H^\prime_4   \:\label{eqc_8}
\ee
Using similar reasoning we can show that 
\be
H_{2n^\prime} =\frac{  H^\prime_{2n^\prime} }{R^\prime_{2n^\prime} \times R^\prime_{2n^\prime} }    \:\label{eqc_9}
\ee
where
\be
R^\prime_{2n^\prime} = 2^{n^\prime} n^\prime!  \:\label{eqc_9b}
\ee
In order to show \cref{eqc_9} one needs to use the fact that the number of non-vanishing terms, $R_{2n^\prime}$, in $H_{2n^\prime}$ is given by
\be
R_{2n^\prime} = (2n^\prime -1 ) R_{2(n^\prime-1)} (2n^\prime -1 ) = [(2n^\prime -1 )!!]^2 \:, \:\label{eqc_10}
\ee
so that the number of independent terms, $r_{2n^\prime}$, associated with each of the two Levi-Civit\`{a} symbols of $H^\prime_{2n^\prime}$ is 
\be
r_{2n^\prime} = (2n^\prime -1)!!  \:.  \:\label{eqc_11}
\ee
Thus, 
\be
R^\prime_{2n^\prime} =  \frac{(2n^\prime)!}{r_{2n^\prime}}  =2^{n^\prime} n^\prime! \:. \label{eqc_12}
\ee
Both \cref{eqc_10,eqc_11} are obtained from combinatorics  arguments.  Using the metric tensor to lower indices in $f^{ij}$ and the relationship \cref{eqc_12} we re-express \cref{eqc_9}
\be
\begin{split}
H_{2n^\prime} = - &\det (g) \frac{1}{(N^\prime - 2n^\prime)!}   \:(2n^\prime -1 )!!^2\:\times  \\   
                & \:^*(\underbrace{f \wedge f \wedge \cdots f}_{n'}) \:\bullet \:^*(\underbrace{f \wedge f \wedge \cdots \wedge f}_{n'})  \:. \:\label{eqc_13}
\end{split}
\ee
The Hodge * operation transforms an $s$-form Q in an $N^\prime$ dimensional space to an $(N^\prime -s)$-form whose components are 
\be
 (^*Q)_{i_{s+1}  \cdots i_{N^\prime}} =   Q^{i_{j_1} \cdots j_{s} } \:
\frac{1}{s!} \:\sqrt{\det |g|} \:\epsilon_{j_1 j_2  \cdots j_{s} i_{s+1}  \cdots i_{N^\prime}} \:,  \:\label{eqc_14}
\ee
and
\be
^*Q \:\bullet \:^*Q \equiv (^*Q)_{i_{s+1}  \cdots i_{N^\prime}} (^*Q)^{i_{s+1}  \cdots i_{N^\prime}} \:\label{eqc_15}
\ee
Note that in \cref{eqc_13}  $\:\underbrace{f \wedge f \wedge \cdots f}_{n'}$ is a $2n^\prime$-form.  Also, in \cref{eqc_13} the minus sign to the right of the equal sign is a consequence of the Minkowski signature of the metric, i.e.\@  $\epsilon^{1 2 \cdots N^\prime}  =-\epsilon_{1 2 \cdots N^\prime}$. For Euclidean signature the minus sign is replaced by a plus sign.  Using properties of the Levi-Civit\`{a} symbol we can show that
\be
H_2 = \det(g) \:\frac{1}{2!} \: f_{ij} f^{ij}  \:, \:\label{eqc_16}
\ee
irrespective of the signature of the metric.  Using \cref{eqc_1b,eqc_13,eqc_16} we obtain
\be
\begin{split}
\det(h) = & \det(g) \:[1 + \frac{1}{2!} \: f_{ij} f^{ij} \mp \sum_{2n^\prime=4}^{2n^\prime=N''}(  \frac{1}{(N^\prime - 2n^\prime)!}   \times  \\  
      &(2n^\prime-1)!!^2 \:^*(\underbrace{f \wedge f \wedge \cdots f}_{n'}) \:\bullet \:^*(\underbrace{f \wedge f \wedge \cdots \wedge f}_{n'})        ) ] \:,   \:\label{eqc_17}
\end{split}
\ee
where
\be
N'' =
\begin{cases}
       N' & \text{if $N'$ is even} \\
			 N'-1 & \text{if N' is odd} 
\end{cases}
\:. \:\label{eqc_17b}
\ee
 The minus (plus) sign corresponds to a metric with Minkowski (Euclidean) signature.   

For the case when $N^\prime = 5$ and the metric has Minkowski signature, \cref{eqc_17} reduces to
\be
\begin{split}
\det(h) =  \det(g) \:[1 + \frac{1}{2!} \: f_{ij} f^{ij} -  &\frac{1}{(5 - 2 \cdot 2 )!}   \times  \\  
      & (3^2 \cdot  1^2 )\:^*(f \wedge f  )_k \:^*(f \wedge f )^k        ] \:.  \:\label{eqc_18}
\end{split}
\ee

\newpage
\bibliography{manuscript}  
\end{document}